\documentclass[useAMS,11pt,twocolumn]{article}
\usepackage{amsmath}
\usepackage[pdftex]{graphicx}

\newcommand{\etendue}{\'etendue }
\newcommand{\etal}{~\it{et al.}\rm~}
\newcommand{\capfont}{\small\textit}

\begin{document}

\title{The dependence of the properties of optical fibres on length}
\author{C. L. Poppett$^{1}$\thanks{E-mail:
c.l.poppett@durham.ac.uk (CLP); j.r.allington-smith@durham.ac.uk (JRAS)} and J. R. Allington-Smith$^{1}$\footnotemark[1]\\
$^{1}$Centre for Advanced Instrumentation (CfAI), Durham University\\}

\date{Accepted for publication, February 2010}

\maketitle
\section{Abstract} 

We investigate the dependence on length of optical fibres used in astronomy, especially the focal ratio degradation (FRD) which places constraints on the performance of fibre-fed spectrographs used for multiplexed spectroscopy. To this end we present a modified version of the FRD model proposed by Carrasco and Parry \cite{Carrasco1994} to quantify the the number of scattering defects within an optical fibre using a single parameter. The model predicts many trends which are seen experimentally, for example, a decrease in FRD as core diameter increases, and also as wavelength increases. However the model also predicts a strong dependence on FRD with length that is not seen experimentally. By adapting the single fibre model to include a second fibre, we can quantify the amount of FRD due to stress caused by the method of termination. By fitting the model to experimental data we find that polishing the fibre causes more stress to be induced in the end of the fibre compared to a simple cleave technique. We estimate that the number of scattering defects caused by polishing is approximately double that produced by cleaving. By placing limits on the end-effect, the model can be used to estimate the residual-length dependence in very long fibres, such as those required for Extremely Large Telescopes (ELTs),  without having to carry out costly experiments. We also use our data to compare different methods of fibre termination.


\section{Introduction}

Focal Ratio Degradation (FRD) is the non-conservation of \etendue  in an optical fibre, resulting in a broadening of the beam at the output.
As noted by Parry \cite{Parry2006}, an increase in \etendue within the instrument without a gain in information content is undesirable because the optical system which has to deal with it is consequently more complex and expensive. As telescopes increase in size, the large optical path lengths caused by the remoteness  of focal stations from gravitationally-invariant instrument platforms and the increased sensitivity to mechanical flexure, mean that optical fibres will continue to be important for the key technique of highly-multiplexed spectroscopy \cite{Allington-Smith2007}. Therefore, it is even more important to be able to predict the performance of optical fibres accurately in order to optimise the instrument system performance. \\

In this paper we propose a modification to a model first proposed by Gloge \cite{Gloge1972} and later adapted by Gambling\etal \cite{Gambling1975} and Carrasco and Parry \cite{Carrasco1994}. The model is described in section \ref{sec:Theory}.  This modification aims to eliminate the dependence on FRD with length which is predicted theoretically but not seen experimentally. By modelling the fibre as two separate fibres with different amounts of scattering defects it can be shown that most of the FRD is caused by the stress frozen in at the fibre ends during end-preparation.

\begin{figure}
\begin{center}
\includegraphics[scale=.5]{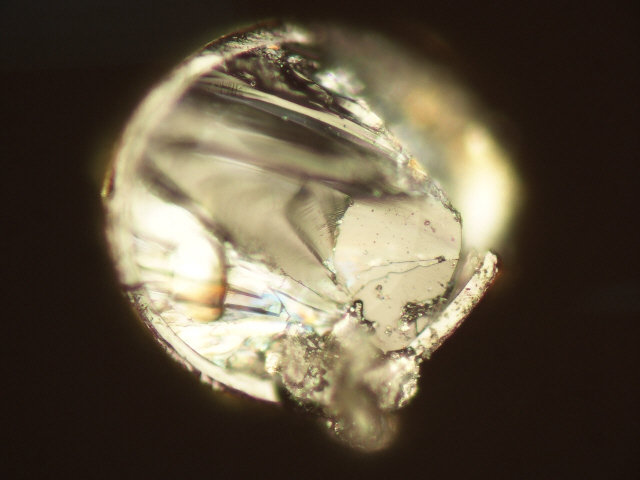}\\
\mbox{\bf (a) } \\
\includegraphics[scale=.5]{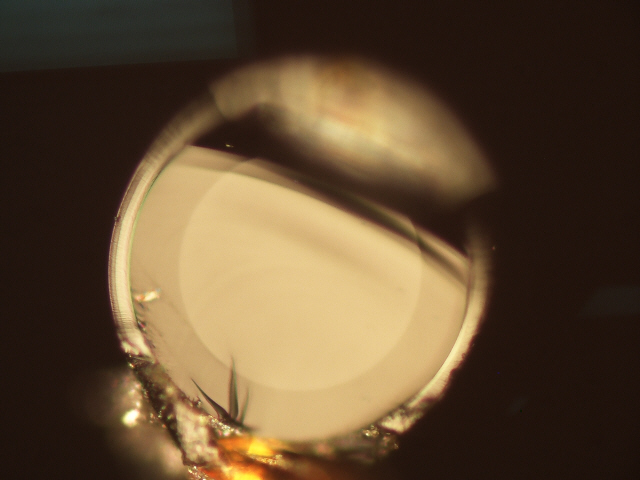} \\
 \mbox{\bf (b)}\\
\includegraphics[scale=.7]{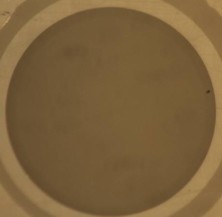} \\
 \mbox{\bf (c)}
\caption{Results of different methods of end-preparation for a typical fibre  with a diameter of $100~\mu m$ (a) unprepared (b) after cleaving  (c) after polishing. In panel (b) the defect from scoring the fibre is clearly visible}
\label{fig:end_prep}
\end{center}
\end{figure}

The need for the fibre ends to be flat, smooth, and perpendicular to the axis of the fibre is well documented, although the method of end preparation varies widely depending on various restrictions. The quickest way to prepare a fibre is cleaving, typically by scoring it  with a diamond pen or tungsten-carbide-edged blade, although a C0$_2$ laser \cite{Kinoshita1979}, or spark erosion \cite{Caspers1976} has also been used. Once the fibre has been scored, it is placed under tension and then bent along the fracture until it snaps. Although this method is relatively quick to perform the results are not always satisfactory and a visual inspection must be performed. In addition to the problem of repeatability, it is extremely difficult to avoid a small defect occurring at the point where the fibre is scored as is shown in panel (a) of figure \ref{fig:end_prep}. The second method is to polish the fibre on progressively finer grades of abrasive paper and finally on a solution of colloidal silica. This method of polishing the fibre is time-consuming, however visual and interferometric inspection of the end normally shows it to be high quality. Haynes\etal \cite{Haynes2008} have shown that the quality of the polish has a large effect on the FRD by comparing fibres with end face roughness of 245~nm  and 6~nm rms.\\

Figure \ref{fig:end_prep} shows the typical finish which can be expected when a fibre is cleaved or polished. However all methods of preparing the ends of the fibre produce stress. The amount of stress and the depth to which it propagates from the end face is currently impossible to measure and methods can only be compared based on the smoothness of the finish as assessed with a microscope or interferometer. The most common rule-of-thumb is that stress propagates to 3 times the depth of the biggest defect on the end face of the fibre although this has never been  proven. It has been shown by Craig\etal \cite{Craig1988} that for the various fibres that they tested, there is no observable difference  in FRD between a properly cleaved fibre and a polished fibre. \\

In section \ref{sec:2fib_model} we modify the Gloge model to include two fibres with different properties in series. This allows us to model the end effect as a short fibre of fixed length with a larger number of scattering defects per unit length, $d_0$, than the main length of fibre. This parameter, $d_0$ has different values depending on the termination method used. \\

It has been shown by Avila\etal \cite{Avila1988} that  FRD is strongly affected by the way in which a fibre is mounted. Avila\etal tested UV grade silica fibres and found that,  at an input focal ratio of  $F_{in}=8$,  $F_{out}$ fell from $F/6.2$ when the fibre was held gently in a clamp to  $F/3.2$ when it was squeezed strongly. This is an important issue because fibres are usually  mounted in a ferrule or holder to protect the tip and allow them to be polished in bulk. When the fibres are mounted in this way they must be glued into position and it is extremely important to test the effects of various adhesives on the resultant FRD  as shown by Oliveria\etal \cite{Oliveira2005}. This model also provides a quantitative measure of these mounting methods which will be independent of the fibre used. \\


\section{Theory Re-visited}
\subsection{Gloge's model}

\label{sec:Theory}
Carrasco and Parry \cite{Carrasco1994} have shown that it is possible to measure the FRD of an optical fibre and characterise its performance using a single parameter, $D$. The model is based on a previous model by Gloge \cite{Gloge1972} who showed that the far-field distribution represents a direct image of the modal power distribution. Gloge developed a partial differential equation to describe the distribution of optical power $P$ in a fibre of length $L$ in terms of the axial angle of incidence $\theta_{in}$, with output angle $\theta_{out}$.\\
\begin{eqnarray}
\frac{\partial P(\theta_{out},\theta_{in})}{\partial L}=&-A\theta_{in}^{2}P(\theta_{out},\theta_{in})+\\ \nonumber
&\frac{D}{\theta_{in}}\frac{\partial}{\partial \theta_{in}}\left (\theta\frac{\partial P(\theta_{out},\theta_{in})}{\partial \theta_{in}} \right ) ,
\label{eqn:Gloge}
\end{eqnarray}
where $A$ is an absorption coefficient and $D$ is a parameter that depends on the constant $d_0$ that characterises microbending:\\
\begin{equation}
D=\left(\frac{\lambda}{2d_fn}\right)^2d_0 ,
\label{eqn:D}
\end{equation}

$\lambda$ is the wavelength of light, $d_f$ the fibre core diameter and $n$ the index of refraction of the core.  Equation \ref{eqn:Gloge} has been solved by Gambling, Payne, and Matsumure \cite{Gambling1975} for the case of a collimated input beam of angle of incidence $\theta_{in}$ :\\

\begin{eqnarray}
\hspace{-1cm}P(\theta_{out},\theta_{in})  = \exp\left \{-\left( \frac{\chi_i+\chi}{2}\right)\left[ \frac{1+{\exp}(-bL)}{1-{\exp}(-bL)}\right] \right \} \\ \nonumber
   \times \left[ \frac{{\exp}(-bL/2)}{1-{\exp}(-bL)}\right]I_o\left[ \frac{(4\chi_i\chi)^{\frac{1}{2}}{\exp}(-bL/2)}{1-{\exp}(-bL)}\right] ,
\label{eqn:Gambling}
\end{eqnarray}

where $\chi=\left( A/D\right)^{1/2}\theta_{in}^2$, $b=4\left(AD\right)^{1/2}$, and $I_0$ is the modified Bessel function of zeroth order.\\

Equation \ref{eqn:Gambling} describes the angular dependence on the output flux in the far field for the case of a collimated input beam at an angle of incidence $\theta_{in}$. It is possible to generalise the solution given by equation \ref{eqn:Gambling} to the case of an input beam with any aperture as follows:\\
\begin{eqnarray}
F(\theta_{out},\theta_0)=&\int_0^{2\pi}\int_0^{\pi}G(\theta_{in},\phi,\theta_{0})\\ \nonumber
&\times P(\theta_{out},\theta_{in}){\sin} \theta_{in}{d}\theta_{in}{d}\phi ,
\end{eqnarray}
where ${\sin}\theta_{in}{d}\theta_{in}{d}\phi={d}\Omega$ is the differential of solid angle, $\phi$ is the azimuthal angle measured around the same axis, and G is a function that represents the input beam. \\

Carrasco and Parry have shown that when the value of $D$ is found for a specific fibre, the results of other experiments where the far field output beam is recorded for input pencil beams of varying focal ratio, can be successfully predicted. By testing a 200 $~\mu$m core fibre Carrasco and Parry found a solution to $D$ which produced a reasonable fit to the experimental data. From their data, the accuracy of optical alignment can be inferred to be ~1 deg since this shifts the predicted output focal ratio by 2.5\% which is the typical size of their experimental errors.\\

As shown by equation \ref{eqn:D} the $D$ parameter depends on the fibre core diameter, $d_f$, the microbending parameter, $d_0$, the wavelength of the injected light, $\lambda$, and the length of the fibre, $L$. From this equation it is clear the $D \propto \lambda^2$, and hence FRD will increase as the wavelength increases. This result has been verified experimentally \cite{Poppett2007}, although the opposite trend has been reported by Murphy\etal \cite{Murphy2008}, who found a weak dependence on FRD with wavelength. However, measurements were made at $F_{in}=3.5$ where FRD is relatively unimportant compared to slower beams where the output beam approaches an asymptotic value. \\

\begin{figure*}
\centering
$\begin{array}{c@{\hspace{. in}}c@{\hspace{. in}}c@{\hspace{.1in}}}
	\includegraphics[scale=0.3]{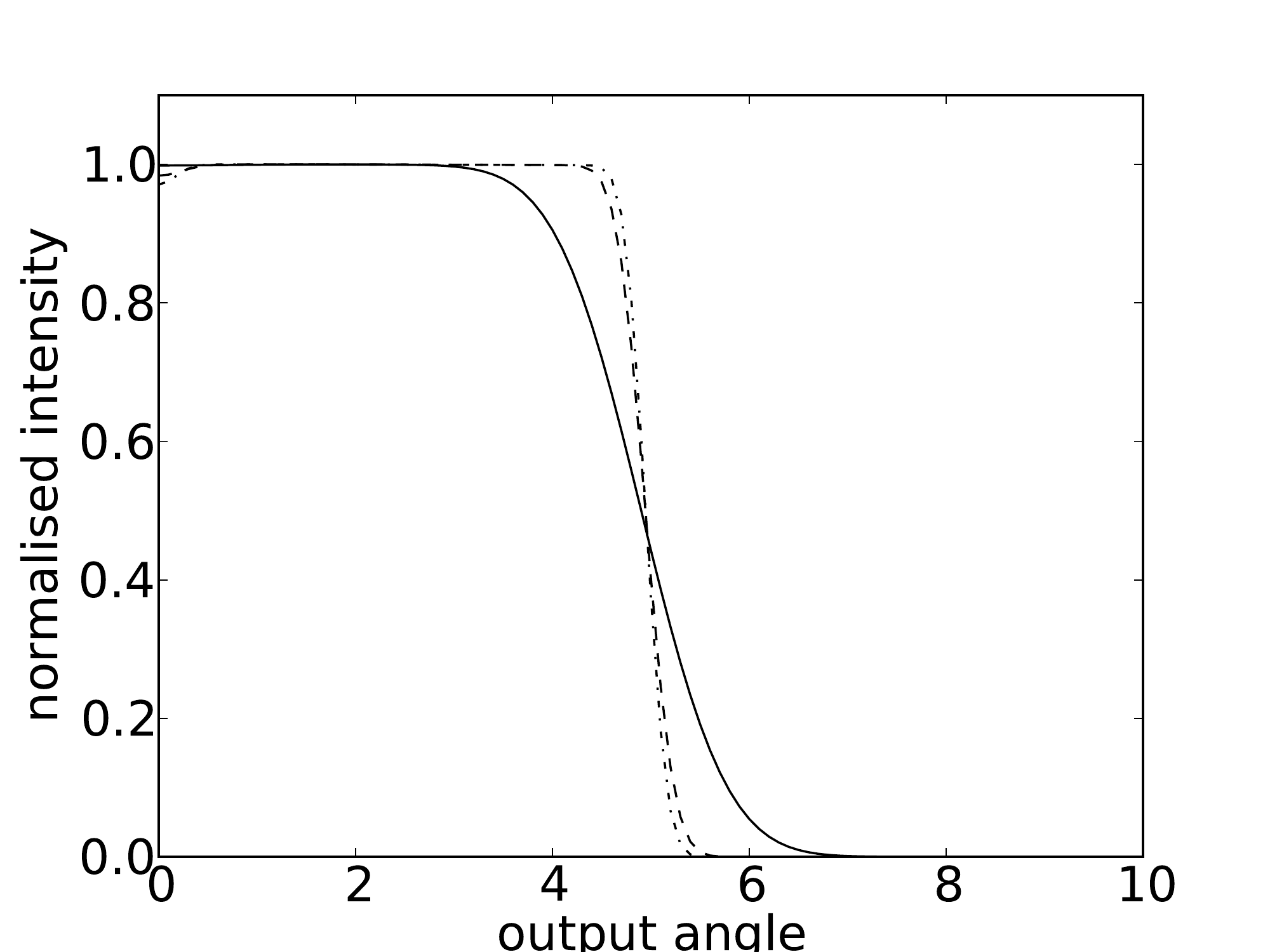} &
	\hspace{0.1cm}
	\includegraphics[scale=0.3]{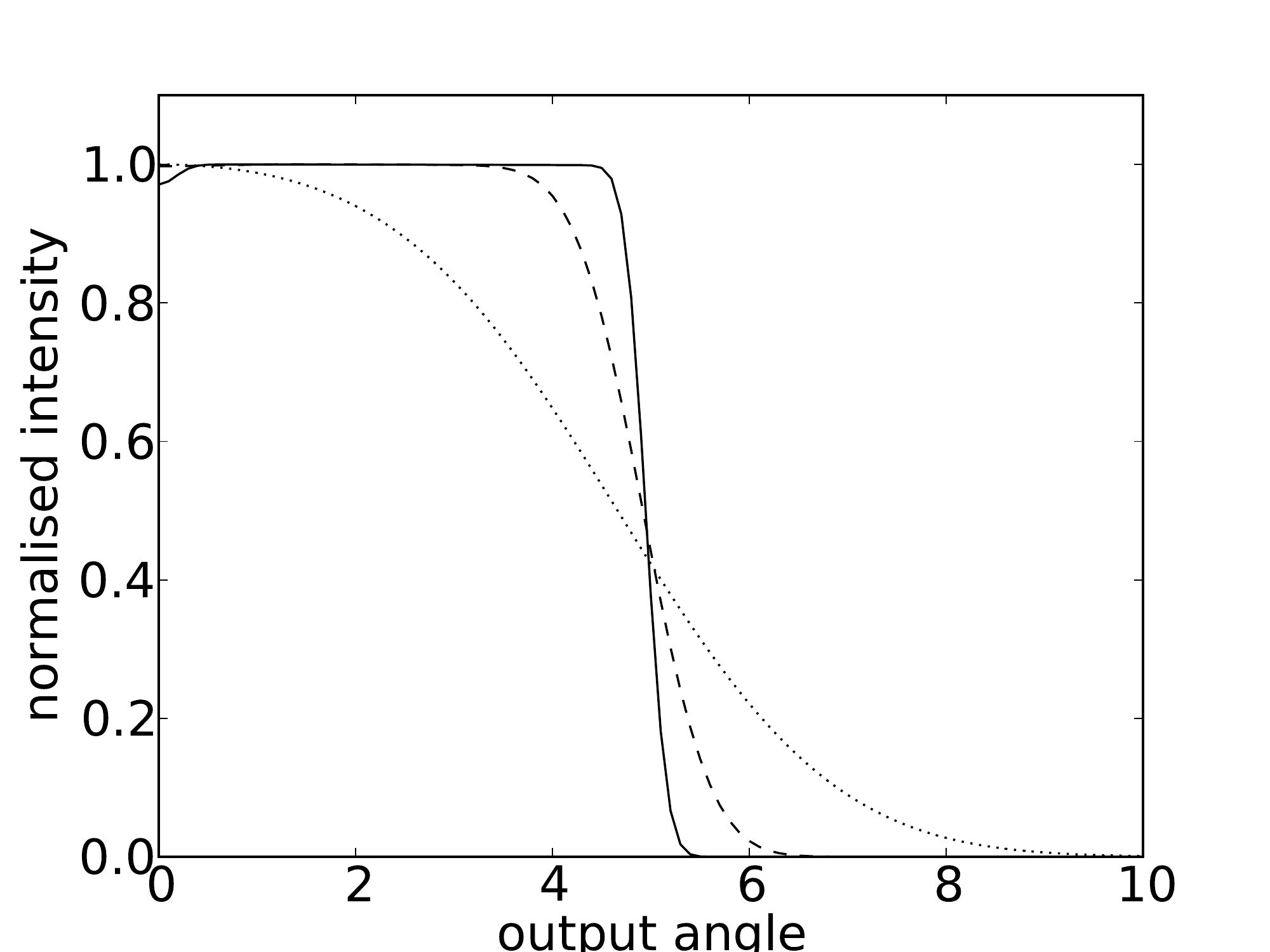}&
	\hspace{0.1cm}
	\includegraphics[scale=0.3]{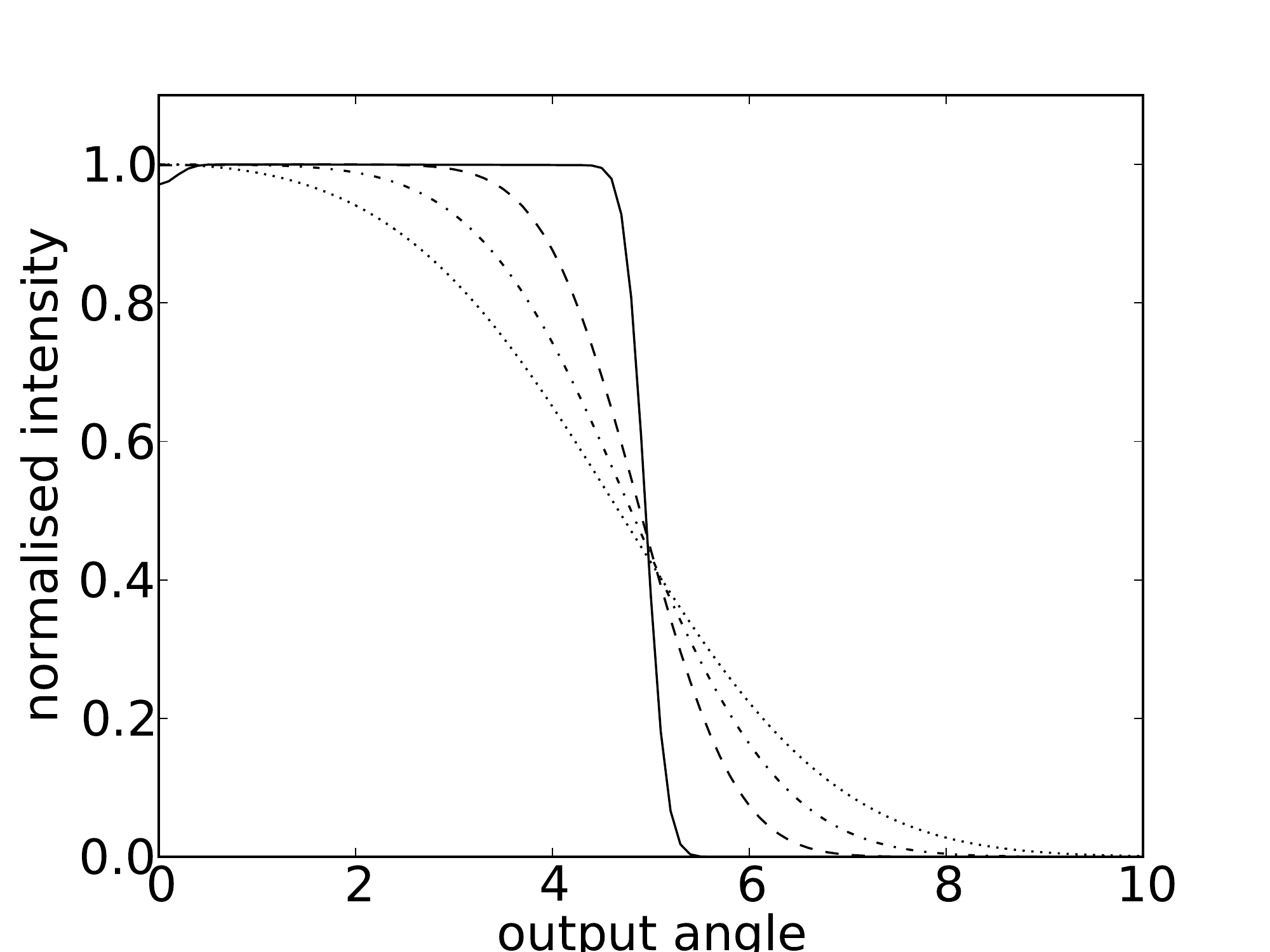}  \\ [0cm]
		\mbox{\bf (a) } & \mbox{\bf (b)}& \mbox{\bf (c) } \\
		\mbox{ $d_f=25 ~\mu m (solid), 75 ~\mu m (dashed),$ } & \mbox{ $d_0=1(solid),20(dashed),$}&\mbox{$L=1~m (solid),10~m(dashed),$}\\\
		\mbox{\bf $100 ~\mu m (dot-dashed)$ } & \mbox{\bf$50(dot-dashed),100(dotted)$}&\mbox{$100m(dot-dashed)$.}\\
	
    \end{array}$
		\caption{\capfont{Normalised theoretical results to show how the output power distribution will be affected by changing the input parameters. All results are for a solid input beam of angle 5$^\circ$ and unless otherwise stated for a fibre of length 1~m, d$_f=100~\mu m$, and $d_0=1$.}\label{fig:theoretical_results}}
\end{figure*}

FRD be quantified conveniently by the FRD ratio $F_{\text{in}}/F_{\text{out}}$.  The second trend predicted by equation \ref{eqn:D} is that $D \propto d_f^{-1}$. Figure \ref{fig:theoretical_results} shows how this model and the two-fibre model described in section \ref{sec:2fib_model}, predict how different parameters will affect the resulting output power distributions. \\

Whilst the trends shown for the fibre core and $d_0$ parameter dependence have been demonstrated experimentally as well as theoretically, the extent to which the length affects the FRD has not. Gambling\etal did report a length-dependence, however this result was for liquid-core fibres. This result has never been reproduced for monolithic fused silica fibres. To confirm this assumption an experiment was set up as described in section \ref{sec:exp}. The following section gives the theoretical description of the method of eliminating this length dependence.

 \subsection{The two-fibre model}
\label{sec:2fib_model}

In order to add a second fibre into the model it is assumed that for a single input ray at angle $\theta_{in}$ to the fibre axis, the output surface brightness at angle $\theta_{out}$ is given by the Gloge model as $F(\theta_{out},\theta_{in})$.\\
For a solid input angle defined by unit surface brightness for $\theta_{in}< \theta_0$, the total output surface brightness is given by:\\
\begin{equation}
F_1(\theta_{out,1},\theta_0)=\int_{0}^{\theta_0}F(\theta_{out,1},\theta_{in,1})2\pi\theta_{in,1}\text{d}\theta_{in,1}\\
\end{equation}

If a second fibre is added to receive light from the first fibre, the equivalent number of single rays it sees at the input is given by:\\
\begin{equation}
\text{d}N{}(\theta_{out,1})\propto F_1(\theta_{out,1},\theta_0) 2\pi\theta_{out,1}\text{d}\theta_{out,1} , \\
\end{equation}

so the output surface brightness in terms of the angle at the exit of the second fibre, $\theta_{out,2}$, is: 

\begin{equation}
\label{eqn:bla}
\begin{split}
F_2(\theta_{out,2},\theta_0) &=\int_{0}^{\infty }F(\theta_{out,2,}\theta_{out,1})\text{d}N{}(\theta_{in,2}) \\
& =\int_{0}^{\infty}F(\theta_{out,2},\theta_{out,1})F_1(\theta_{out,1},\theta_0)\\
&\hspace{1cm}\times 2\pi\theta_{out,1}\text{d}\theta_{out,1}
\end{split}
\end{equation}

The coupling between fibres is achieved by setting $\theta_{in,2}=\theta_{out,1}$.\\

\begin{figure}
\centering
	\includegraphics[scale=0.3]{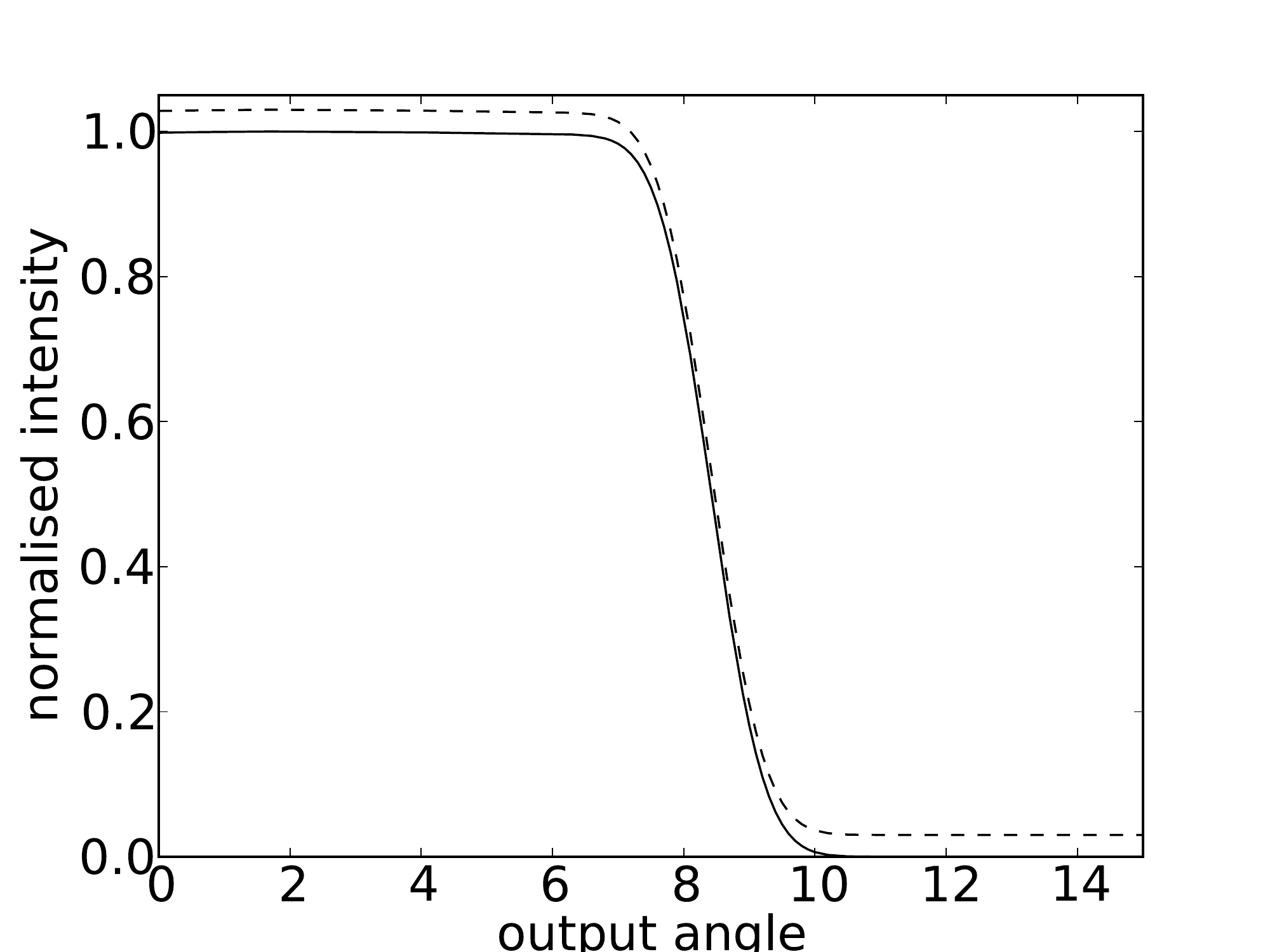} \\
	\mbox{\bf (a) }\\
	\includegraphics[scale=0.3]{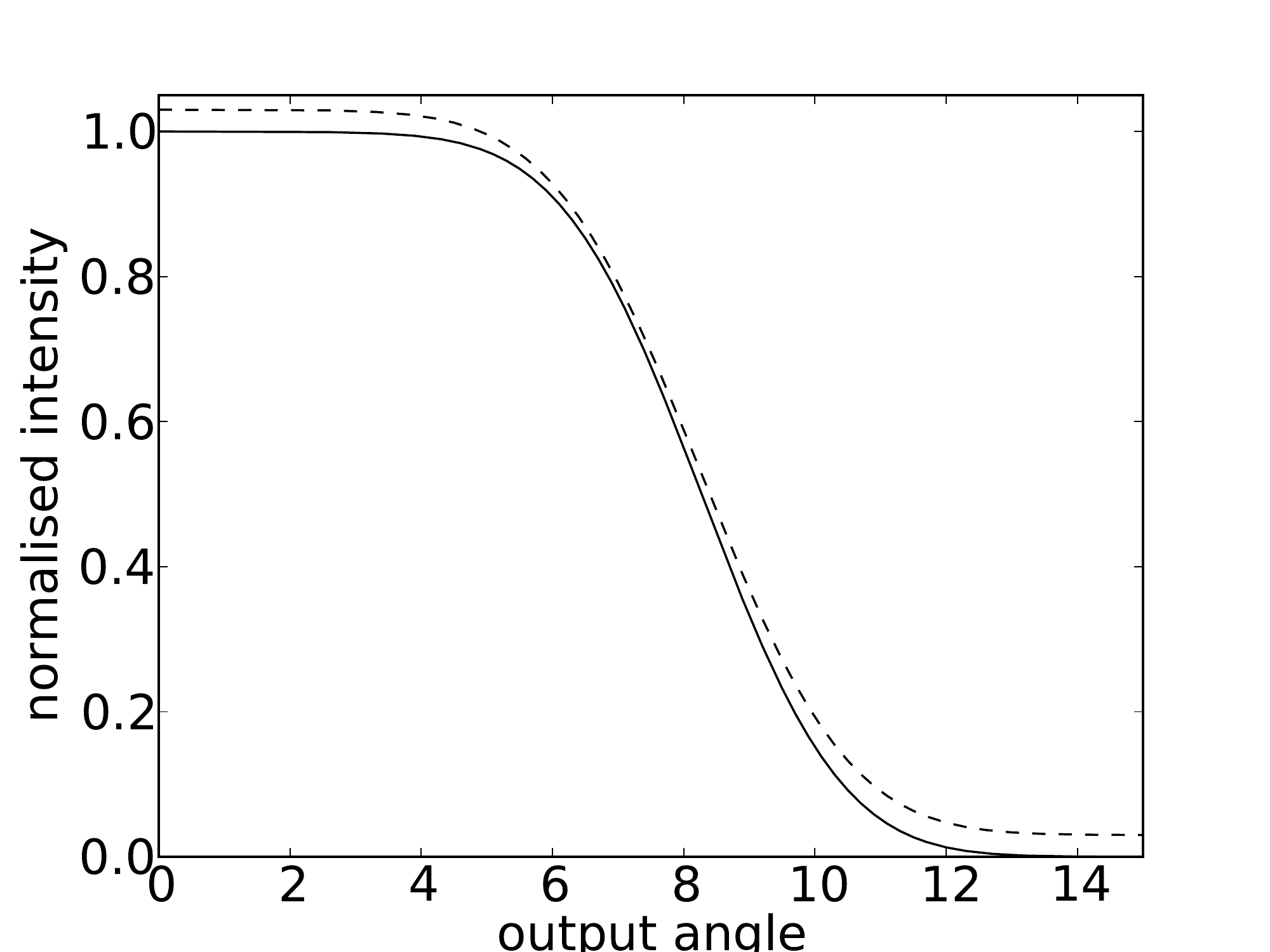}  \\ 
	\mbox{\bf (b)}
 
		\caption{\capfont{output power distributions predicted by the one fibre (dashed) and two fibre models (solid, with an offset of 0.03 in order to be able to show both curves).\label{fig:model_comparisons}}}
\end{figure}

Figure \ref{fig:model_comparisons} shows curves for the output power distribution produced by both the one-fibre and two-fibre model.  Since we can't determine $d_{0,2}$ and $L_2$ independently, we use the product $S = d_{0,2} L_2$ to characterise the second fibre. This parameter  essentially quantifies the total number of scattering centres (microbends ) present in the second fibre. The value of $S$ was  chosen to produce 
the same amount of FRD at the asymptotic value as the one-fibre model. This figure  shows that the two-fibre model agrees well with the predictions made by the one-fibre model (itself rigorously tested against experimental data here and elsewhere). Indeed the difference in the 95\% enclosed energy values  of the one- and two-fibre models differ by the same amount as the experimental error bars (see below). We use the angle which encloses 95\% of the energy to define the output focal ratio since this reduces sensitivity to stray light in the experimental setup.

\section{Experimental results}
\label{sec:exp}

\begin{figure*}
\begin{center}
\includegraphics[scale=0.5]{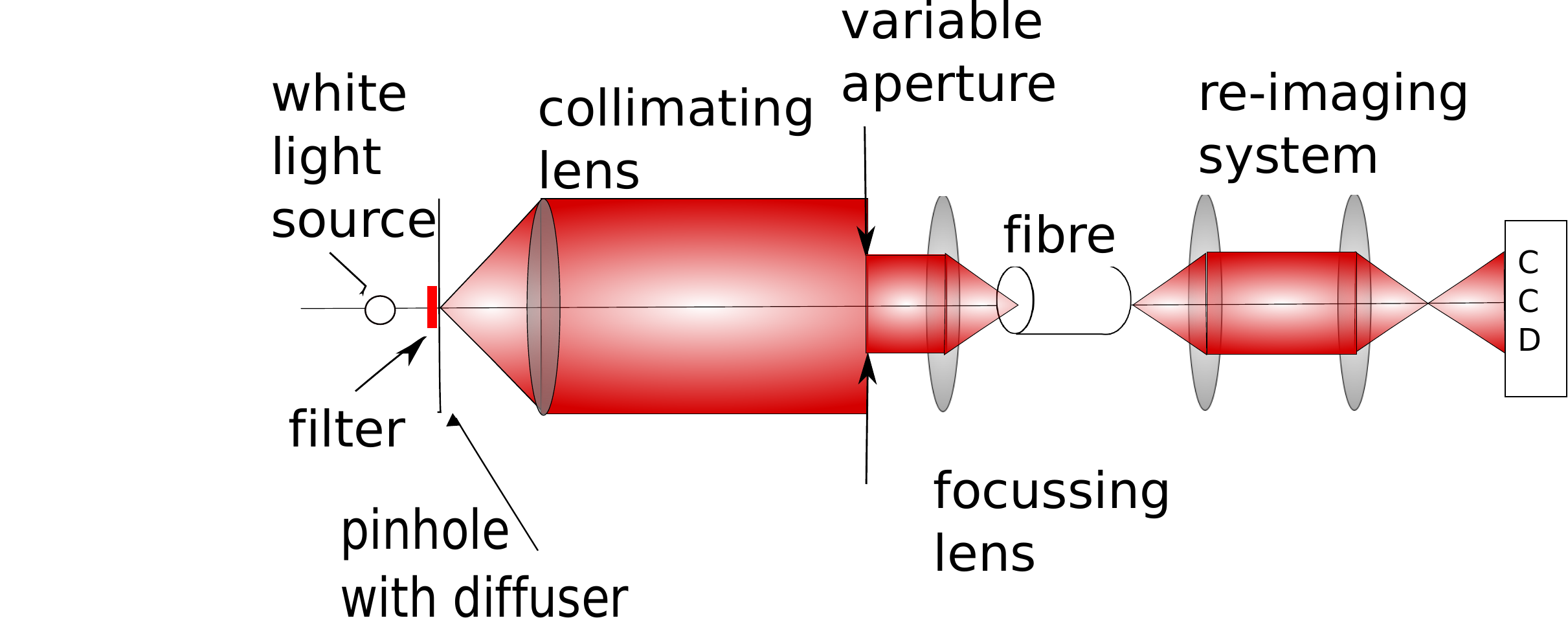}
\caption{Experimental setup. The adjustable iris was used to give a range of input focal ratios and a beam splitter was used in order to provide an image of the input light on the end face of the fibre.}
\label{fig:setup}
\end{center}
\end{figure*}

To test the experimental dependence on FRD with length, a fibre was illuminated with a solid input beam of varying $\theta_{in}$ as shown by figure \ref{fig:setup}. The fibre used was a typical fibre made by polymicro ( FIP100110125) with a core diameter of 100~$\mu m$, which has similar characteristics to fibres used in astronomical instruments. Input beams ranging between f/1.7 and f/18.9 (i.e. cone half-angles $16^\circ-1.5^\circ$) were used to investigate how the FRD evolved with input angle.\\

\begin{figure*}
\begin{center}
\includegraphics[scale=0.5]{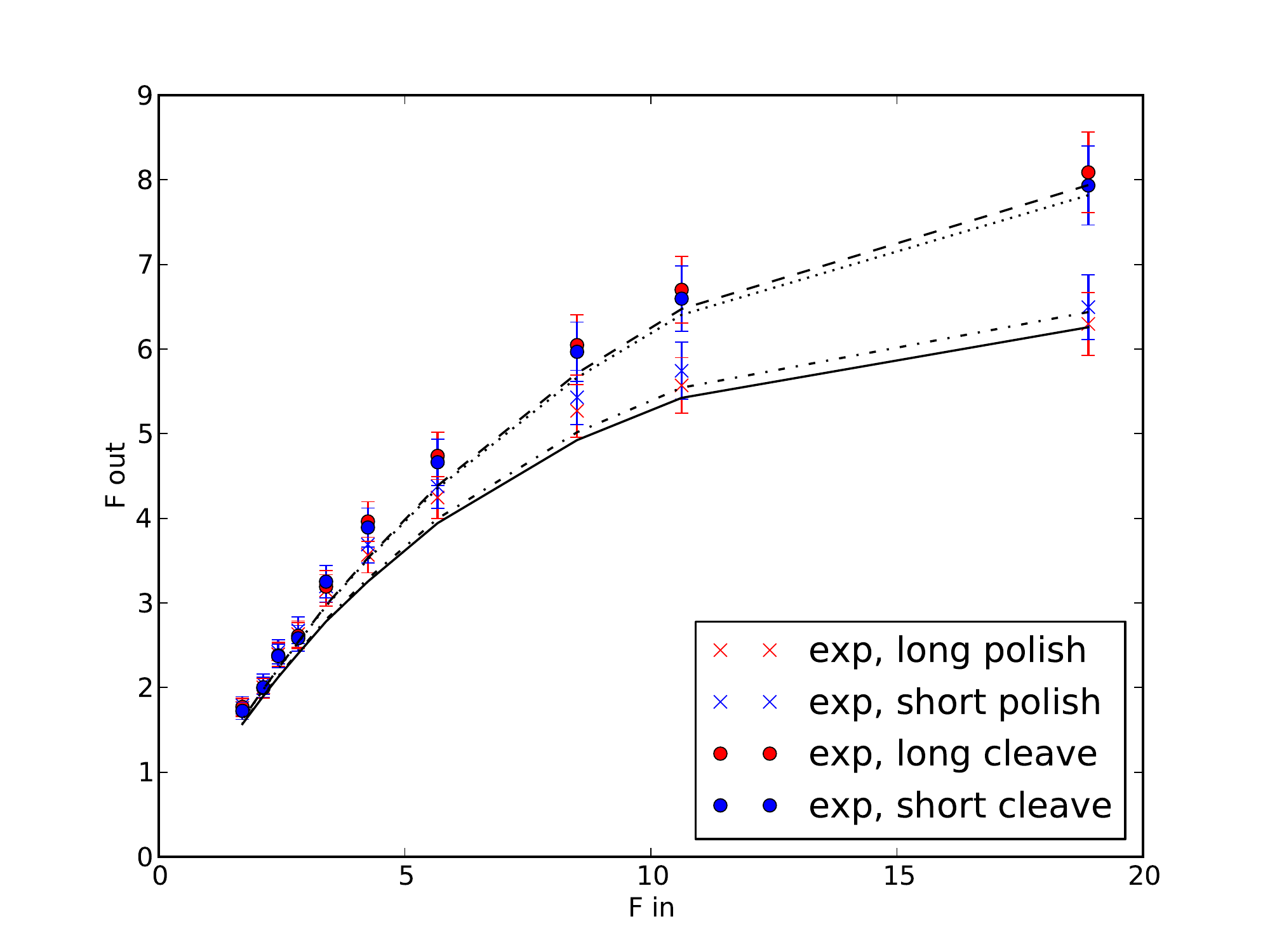}
\caption{FRD curves where experimental results are shown by crosses and theoretical results produced by the original one-fibre model are given by a smooth line}
\label{fig:exp_results}
\end{center}
\end{figure*}

Tests were taken for a case of a cleaved fibre of length 10~m immersed via index-matching gel to a smooth flat glass plate and for the same fibre after it had been polished to a 1$~\mu m$ optical finish. The FRD curve presented in figure \ref{fig:exp_results} show that the fibre has a fairly good FRD performance with an FRD ratio of around 1.1 at $F_{in}=4.5$.  The rationale for testing a cleaved fibre with index matching gel was to make sure that we isolated the effect of stress from any scattering caused by imperfections in the topography of the end face. The use of index-matching gel for the cleaved fibre should eliminate the effect of surface roughness and the smoothness of the finish  would  do the same for the polished fibre. Furthermore, we expected that cleaving would result in relatively low stress and therefore provide a baseline against which the results of the polishing could be compared.

The test was repeated after the fibre had been cut to a length of 1~m for both the cleaved polished case. These two lengths of fibre will be referred to as 'long' and 'short' for the remainder of this report. \\

The measured FRD for all scenarios under which the fibre was tested are also shown in figure \ref{fig:exp_results}. From this, the $D$ parameter for the fibre has been estimated in each of the specific cases (Table \ref{tbl:D_values}). Theoretical predictions using these parameters are shown in figure \ref{fig:exp_results}. From this we estimate that the difference in FRD due to the different fibre  lengths  for the cleaved fibre at the asymptotic value was $0.15 \pm 0.7$, and hence is clear that within our experimental uncertainty, no length dependence is observed.

\begin{table*}
\caption{Parameters need to produce theoretical curves which match experimental results for the fibre preparation methods.}
\begin{center}
\begin{tabular}{ccccc}
preparation&\hspace{10pt} fibre &\hspace{10pt} FRD ratio &\hspace{10pt}$d_0$ &\hspace{10pt}D \\
method&\hspace{10pt}length&\hspace{10pt}(at $F_{in}=4.5$) &\hspace{10pt}($m^{-1}$)&\hspace{10pt}($m^{-1}$)\\
\hline
\hline
cleave & \hspace{10pt}long &\hspace{10pt} 1.07&\hspace{10pt} 5.0&\hspace{10pt}$2.1\times 10^{-5}$\\
\hline
cleave & \hspace{10pt}short &\hspace{10pt} 1.09  &\hspace{10pt} 52&\hspace{10pt}$2\times 10^{-4}$\\
\hline
polish &\hspace{10pt}long &\hspace{10pt} 1.19 &\hspace{10pt} 9.1&\hspace{10pt}$4\times 10^{-5}$\\
\hline
polish & \hspace{10pt}short &\hspace{10pt} 1.15 &\hspace{10pt} 85&\hspace{10pt}$3.7\times 10^{-4}$\\
\end{tabular}
\end{center}
\label{tbl:D_values}
\end{table*}%

\section{Discussion of results}
We now use  the experimental results to eliminate the length dependency and compare stress induced by different end preparation techniques. To this end the model can be run for various combinations of $d_{0,1}$ and $S$ (where $S=d_{0,2}L_2$) to find a single set of values which predict the experimentally-observed FRD curves.  In order to find this set of values, a number of constraints can be applied to the two-fibre model, such that we require:
\begin{enumerate}
\item{$d_{0,1}, S$ combinations where the difference in asymptotic $F_{\text{out}}$ for the long and short fibres is less than the experimental errors;}
\item{$d_{0,1}, S$ combinations which give an $F_{\text{out}}$ which agrees with the experimental data at the asymptote for both the cleaved and polished fibres ;}
\item{$d_{0,1}$ value less than that determined from  the one-fibre model, and the same  for both the cleaved and polished fibres.}
\end{enumerate}

\begin{figure*}
\begin{center}
\includegraphics[scale=0.3]{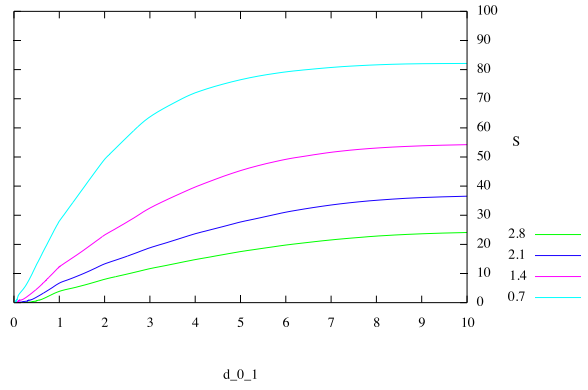}
\includegraphics[scale=0.3]{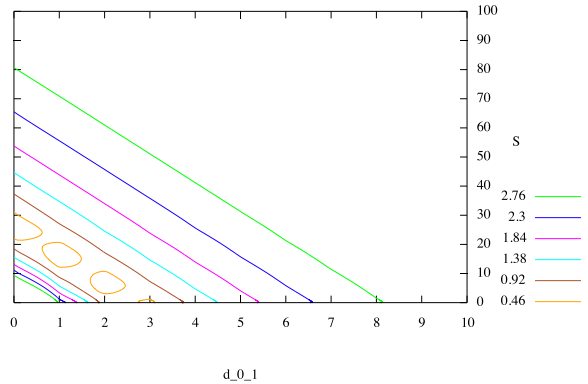}\\
\mbox{\bf (a) \rm{difference in asymptotic $F_{out}$ between L=1~m } } \hspace{1cm} \mbox{\bf (b) \rm difference in asymptotic $F_{out}$ between experimental }\\
\hspace{-1cm}\mbox{and L=10~m with contours shown in increments of 0.7} \hspace{0.7cm}\mbox{ and theoretical results for the cleaved fibre}\\
\vspace{-2.5cm}
\includegraphics[scale=0.5]{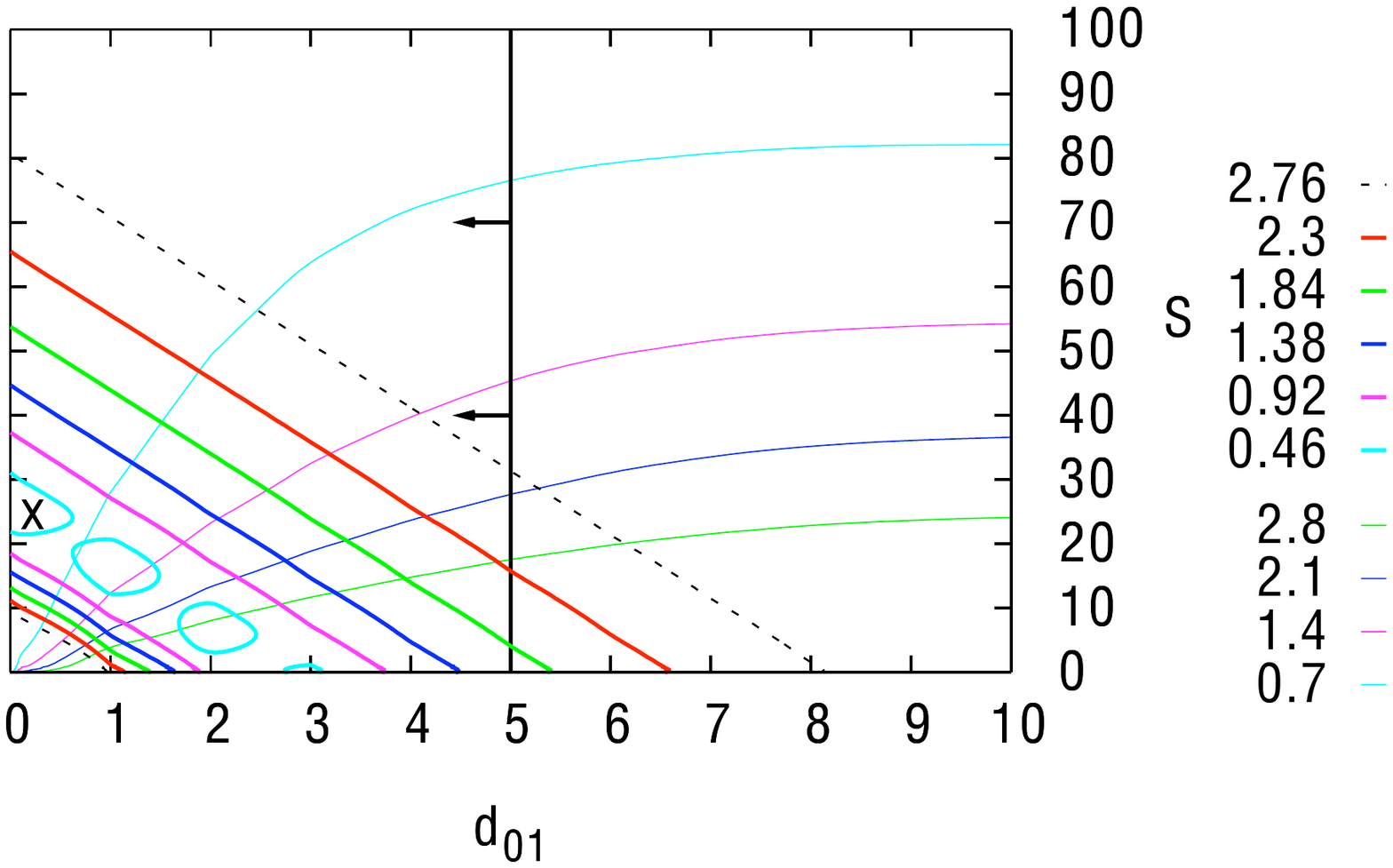}\\
\vspace{-2cm}
\mbox{\bf (c) }
\caption{The two fibre model was run for various combinations of $d_{0,1}$ and $S$ (where $S=d_{0,2}L_2$) . Panel (a) shows the difference in asymptotic $F_{out}$ between L=1~m and L=10~m with contours shown in increments of 0.7. Panel (b) shows the difference in asymptotic $F_{out}$ between experimental and theoretical results for the cleaved fibre. Panel (c) shows how these graphs can be combined to find a single solution, marked by the large cross, for  $d_{0,1}$ and $S$ which satisfies experimental data. The black vertical line in panel (c) shows the $d_{0,1}$ parameter predicted by the one fibre model for the cleaved fibre.}
\label{fig:contour}
\end{center}
\end{figure*}

As is shown in section \ref{sec:exp}, the length dependence was less than $F/0.15 \pm 0.67$ for the cleaved fibre. The contour plot shown in panel (a) of figure \ref{fig:contour} defines the upper limit on allowable $\Delta F$, and all values within the 0.7 contour region give acceptable solutions for specific values of $d_{0,1}$ and $S$ . Secondly,  the asymptotic $F_{\text{out}}$ must be the same as the experimental data within the error bars. For the case of the cleaved fibre, the error in $F_{out}$ was determined to be 0.46 and this contour is shown in panel (b) of figure \ref{fig:contour}. All values within this 0.46 contour region give acceptable solutions for specific $d_{0,1}$ and $S$ . The final panel in figure \ref{fig:contour} shows how these constraints combine, and a solution for the cleaved fibre is shown by a cross at $d_{0,1}=0.1~m^{-1}$ and $S=25$. Using these constraints for the polished fibre produces a solution at $d_{0,1}=0.1~m^{-1}$ and $S=48$. \\

\begin{figure*}
\begin{center}
\includegraphics[scale=0.4]{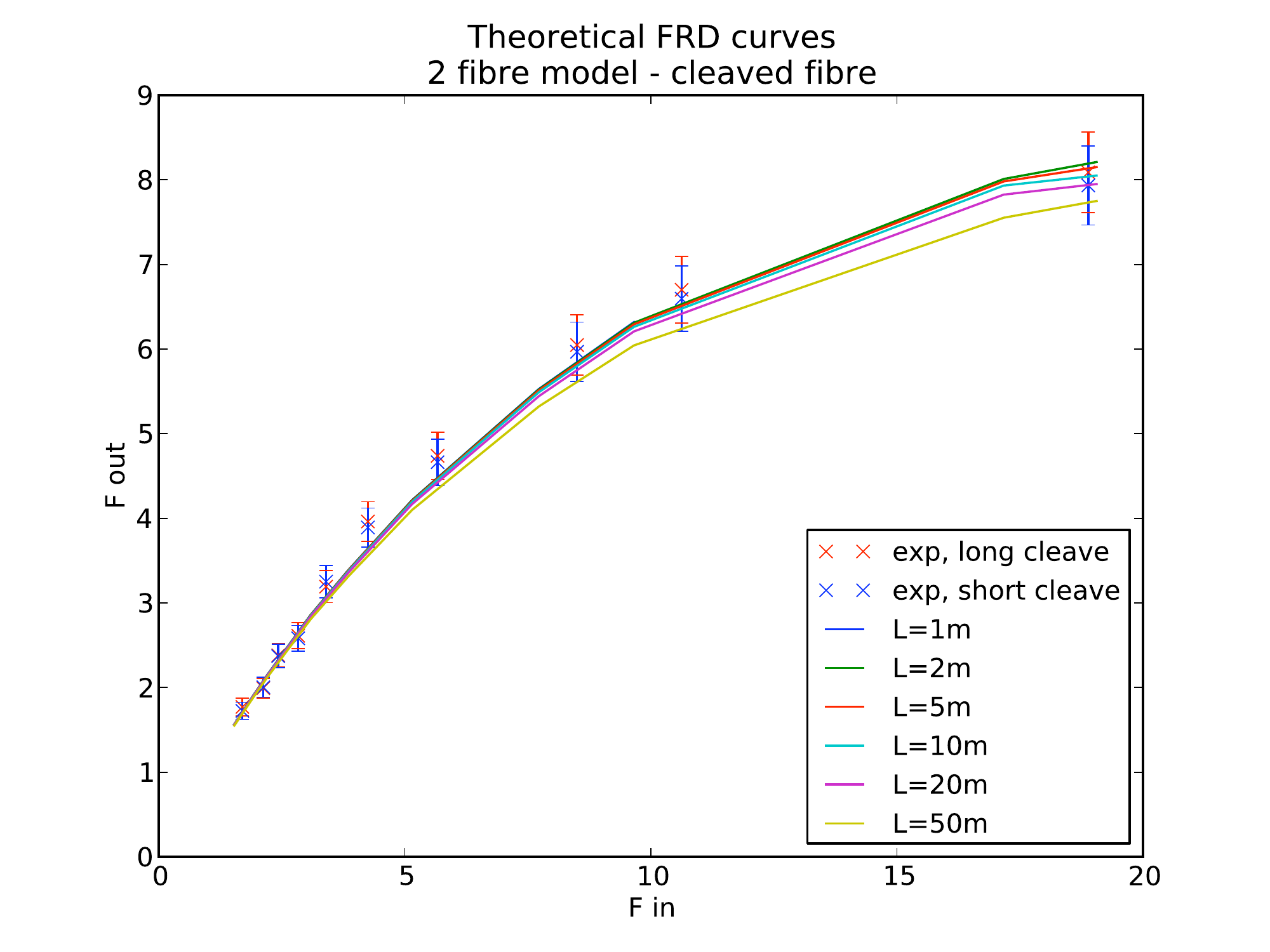}
\includegraphics[scale=0.4]{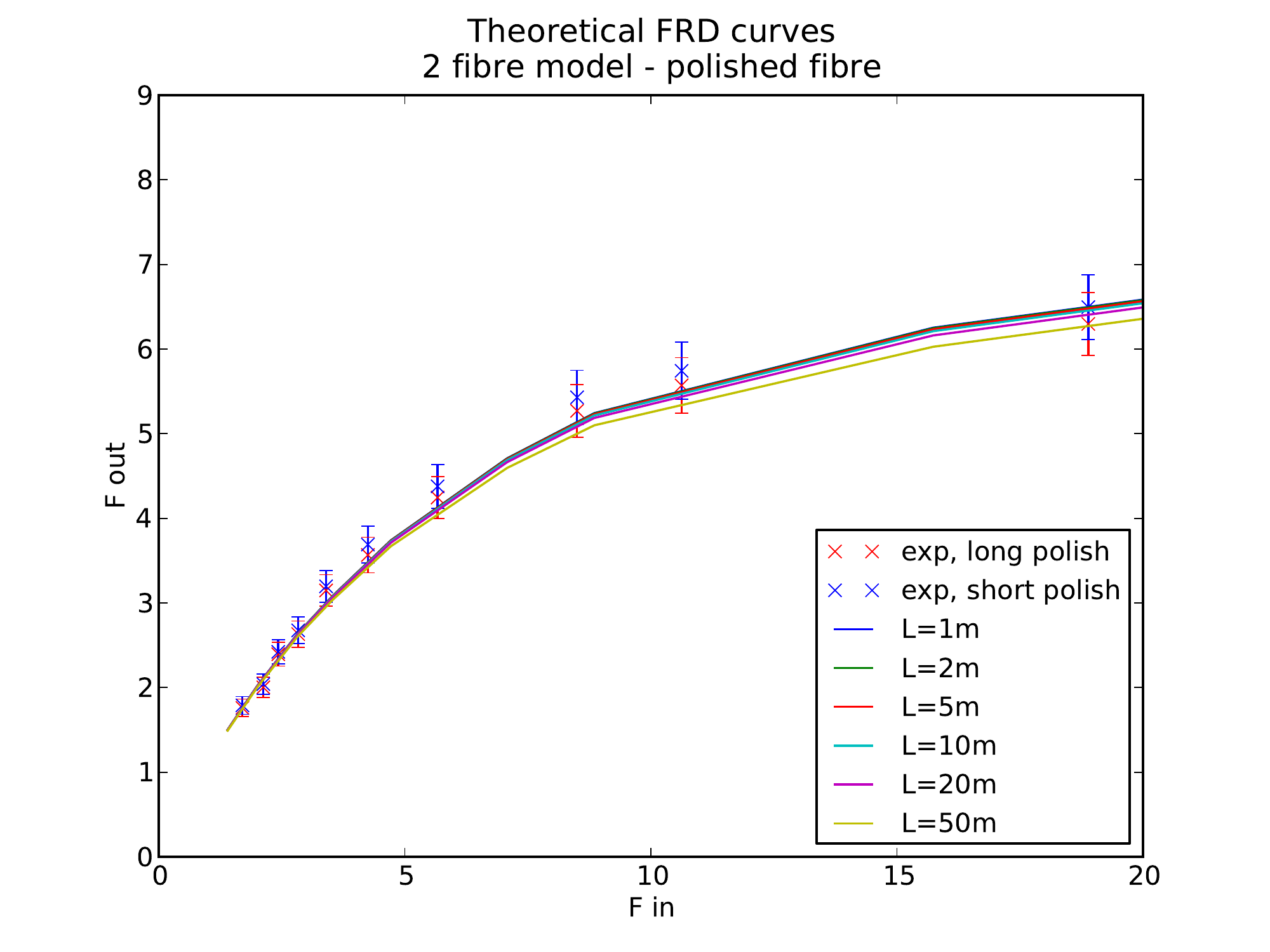}

\caption{2 fib model results with $d_{0,1}=0.1~m^{-1}$ and $d_{0,2}L_2=25$ for the cleaved fibre and  $d_{0,2}L_2=48$ for the polished fibre. It is clear that within the errors the length dependence of the original model has been removed.}
\label{fig:2_fib_cleave}
\end{center}
\end{figure*}

Figure \ref{fig:2_fib_cleave} shows the FRD curves for various lengths of fibre using these parameters. From this figure it is clear that we have removed the length dependence of the model within the lengths for which it was constrained. The small discrepancies between the theoretical and experimental results arise due to misalignments within the experiment and the fact that we have assumed $G$ was a perfect step function rather than measuring it explicitly.\\

The second objective of the two fibre model is to provide a method of comparing stress induced by various end preparation techniques. Comparison of the parameter, $S$, between the polished and cleaved cases gives a quantitative assessment of the induced stress. The relative values show that there are twice as many microbends per unit length when polishing the fibre, compared to when it is cleaved. The high absolute values of $S$ make a very strong argument for taking great care to eliminate stress in the end-preparation process.

Currently, there is no method of determining the depth to which stress in the end of the fibre extends. The rule-of-thumb that the stress propagates to 3 times the depth of the largest defect on the end face of the fibre implies that the stress only affects a few $\mu$m of the length of the fibre and forces $d_{0,2}$ to be extremely large. Thus, assuming that the defect size is determined by the smallest grit size, 
 implies $L_2=3~\mu m$ and $d_{0,2}= 8.3\times 10^{6} ~m^{-1}$ and $d_{0,2}=16.0\times 10^{6} ~m^{-1}$ for the polished and cleaved fibres respectively. 

\section{Conclusions}
We have adapted the power distribution model described by Gloge and produced a new model which will eliminate the length dependence on FRD in order to provide a better agreement with experimental data. This model has been tested against the original one-fibre model, and with experimental data, and shown to be in good agreement. As a result of trying to quantify the amount of stress within the end of the fibre it has been shown that the end-effect is extremely powerful. If we assume that the $d_{0,2}$ parameter characterizes the FRD induced in the end of the fibre then it is possible to compare how the FRD produced in the majority of the length of the fibre is affected. In the case of the cleaved fibre  the $d_0$ parameter (which quantifies the amount of microbending) calculated using the one fibre model was reduced from $5~m^{-1}$ for a $1~m$ fibre and $52~m^{-1}$ for a $10~m$ fibre to $0.1~m^{-1}$ independent of length using the two-fibre model. This significant reduction is testament to the strength  of the end effect and clearly shows how much FRD is generated in the end of the fibre. 

It was initially assumed that cleaving the fibre would produce very little stress, however this was not shown to be the case. This model is therefore extremely useful when comparing various end preparation techniques. For example, our results show that cleaving the fibre halves the strength of  the end effect compared with polishing. The model can also provide an upper limit on the FRD performance of any length of fibre and will therefore prove to be very useful when building instruments for Extremely Large Telescopes. 


\bibliography{FRD_end_effects}
\bibliographystyle{unsrt}

\end{document}